\documentclass[preprint]{elsarticle}
\pdfoutput=1

\usepackage{lineno,hyperref}
\modulolinenumbers[250]

\journal{Nuclear Physics B (Elsevier)}

\usepackage{amsfonts}
\usepackage{mathrsfs}
\usepackage{amsmath}
\usepackage{xcolor}
\usepackage{amssymb}
\usepackage{bm}
\usepackage{feynmp}
\usepackage{extarrows}
\usepackage{slashed}
\usepackage{graphicx}
\usepackage{subfigure}

\renewcommand{\thefootnote}{\fnsymbol{footnote}}

\newcommand{\fr}[2]{\mbox{$\frac{\,{#1}\,}{#2}$}}

\def\bge{\begin{equation}}
\def\ede{\end{equation}}
\def\bga{\begin{aligned}}
\def\eda{\end{aligned}}
\newcommand{\beq}{\begin{equation}}
\newcommand{\eeq}{\end{equation}}
\newcommand{\bq}{\begin{equation}}
\newcommand{\eq}{\end{equation}}
\newcommand{\ba}{\begin{array}}
\newcommand{\ea}{\end{array}}
\newcommand{\beqa}{\begin{eqnarray}}
\newcommand{\eeqa}{\end{eqnarray}}
\newcommand{\beqs}{\begin{subequations}}
\newcommand{\eeqs}{\end{subequations}}

\def\nn{\nonumber}
\def\non{\nonumber}

\def\({\left(}
\def\){\right)}

\def\End{\end{document}}

\def\leqq{\leqslant}

\def\to{\rightarrow}

\def\End{\end{document}}

\begin{document}

\begin{frontmatter}

\title{{\bf Single-Valued Hamiltonian via Legendre-Fenchel Transformation
       and Time Translation Symmetry}}

\author{{\sc Huan-Hang Chi}\,$^{a,b,c,}$\footnote{Email: hhchi@stanford.edu}
        ~and~
        {\sc Hong-Jian He}\,$^{b,c,d,}$\footnote{Email: hjhe@tsinghua.edu.cn}
}

\address[a]{Physics Department, Stanford University, Stanford, California 94305, USA\\[1mm] }
\address[b]{Institute of Modern Physics and Center for High Energy Physics, \\
            Tsinghua University, Beijing 100084, China\\[1mm] }
\address[c]{Physics Department, Tsinghua University, Beijing 100084, China\\[1mm] }
\address[d]{Center for High Energy Physics, Peking University, Beijing 100871, China\\[18mm] }

\begin{abstract}
Under conventional Legendre transformation,
systems with a non-convex Lagrangian will result in a multi-valued Hamiltonian
as a function of conjugate momentum. This causes problems such as non-unitary
time evolution of quantum state and non-determined motion of classical particles, and
is physically unacceptable. In this work, we propose a new construction of single-valued Hamiltonian
by applying Legendre-Fenchel transformation, which is a mathematically rigorous
generalization of conventional Legendre transformation, valid for non-convex Lagrangian systems,
but not yet widely known to the physics community. With the new single-valued Hamiltonian,
we study spontaneous breaking of time translation symmetry and derive its vacuum state.
Applications to theories of cosmology and gravitation are discussed.
\end{abstract}

\begin{keyword}
Non-convex Lagrangian, Legendre-Fenchel Transformation, Single-Valued Hamiltonian, Time Translation Symmetry
\\[2mm]
Nucl.\ Phys.\ B (2014), in Press [arXiv:1310.3769 [quant-ph]].
\end{keyword}

\end{frontmatter}

\linenumbers

\renewcommand{\thefootnote}{\arabic{footnote}}
\setcounter{footnote}{0}

\newpage

\baselineskip 16.5pt

\section{Introduction}

Conventional physical systems are described by Lagrangians which are
convex functions of velocity. Thus, one can derive unique single-valued
Hamiltonians by applying the Legendre transformation.
This leads to well-defined general formalism of classical and quantum theories.
Their Lagrangians are quadratic in velocities with positive coefficient,
and are thus convex functions.
But, physical systems with non-convex Lagrangians are also very interesting,
since they are important for studying spontaneous breaking
of time translation symmetry \cite{FW-3,FW-1,PRA} and are widely applied
to theories of cosmology and gravitation \cite{Cos1,Cos2,Cos2b,GR-topo,GR-topo2,GR-new}.

Recently, Shapere and Wilczek considered interesting models with non-convex Lagrangians
in velocity \cite{FW-3,FW-1}.
For the purpose of demonstration,
let us consider a simple model \cite{PRA,FW-3,FW-1},
\begin{equation}
\label{original_L}
L \,=\, \frac{1}{4}\dot{\phi}^4-\frac{\kappa}{2}\dot{\phi}^2 \,,
\end{equation}
to demonstrate the essential idea.
For the nontrivial case of $\,\kappa > 0\,$,\,
the Lagrangian is a non-convex function of velocity.
Thus, the conjugate momentum
\begin{equation}
\label{momentum}
p \,=\, \frac{\partial L}{\partial \dot{\phi}}
  \,=\, \dot{\phi}^3-\kappa\dot{\phi}
  \,\equiv\, f(\dot{\phi}) \,,
\end{equation}
is not monotonic in velocity, where the function $\,f(\dot{\phi})\,$
stands for the Legendre map.
Then, making the conventional Legendre transformation gives the corresponding
Hamiltonian as a function of velocity,
\begin{equation}
\label{muti-H}
H \,=\, \frac{3}{4}\dot{\phi}^4-\frac{\kappa}{2}\dot{\phi}^2 \,,
\end{equation}
which is a multi-valued function (with cusps) in conjugate momentum $\,p\,$,\,
since each given $\,p\,$ corresponds to one or three values of $\,\dot{\phi}\,$,\,
as shown in Eq.\,(\ref{momentum}).
\begin{figure}
   \begin{center}
     \includegraphics[width=0.65\textwidth]{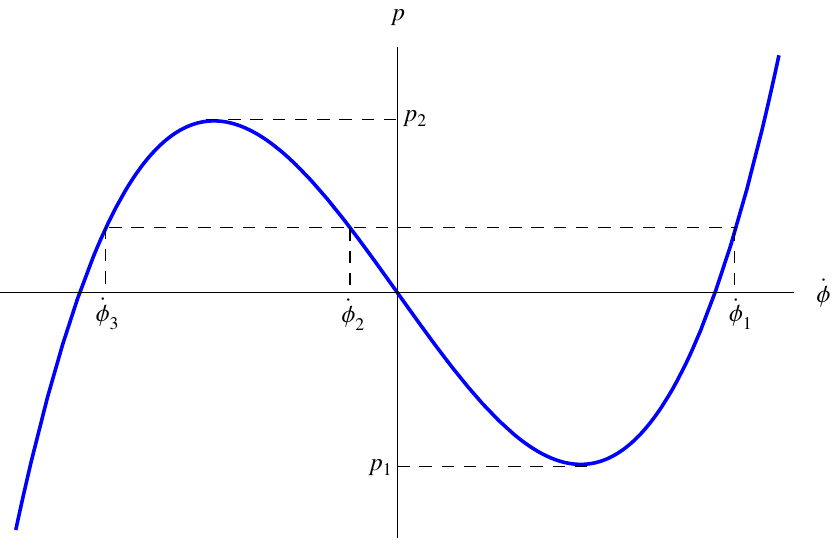}
     \vspace*{-3mm}
   \end{center}
\caption{The Legendre mapping function $\,f\,$ from $\,\dot{\phi}\,$ to $\,p\,$,\,
for the non-convex Lagrangian (\ref{original_L}) with $\,\kappa > 0\,$.}
\label{p-velocity-fig}
\end{figure}

A multi-valued Hamiltonian will make the evolution of quantum state ill-defined.
Since at any moment, for a given $\,\phi\,$ and $\,p\,$,\,
one does not know which ``branch'' of the multi-valued Hamiltonian
could be used to generate the evolution of quantum state.
This means that the time evolution of a quantum state may arise from any branch
of the naive multivalued Hamiltonian and switch from one branch to another\,\cite{PRA},
which would cause nonunitary evolution and thus be physically unacceptable\,\cite{PRA}.

For classical motion, similar reasoning also applies to the Hamiltonian formulation.
We can recast this problem by using the Lagrangian (\ref{original_L}).
The equation of motion from the Lagrangian (\ref{original_L}) is given by
\begin{equation}
\label{EOM1}
\frac{d}{dt}p(\dot{\phi}) \,=\, 0 \,,
\end{equation}
which requires
\begin{equation}
\label{EOM2}
p(\dot{\phi}) \,=\, \dot{\phi}^3-\kappa\dot{\phi}
\,\equiv\, f(\dot{\phi}) \,=\, p_0^{} \,,
\end{equation}
to be constant in time,
where $\,p_0^{}\,$ is given by the initial condition.
[Since $\,\dot\phi\,$ is not continuous and thus non-differentiable with $t$ \cite{PRA},
$\,\ddot\phi\,$ is not defined.
So we will use the integral version \eqref{EOM2}, instead of \eqref{EOM1}.]
As mentioned earlier, for some $\,p_0^{}\,$ values such as
$\,p_1^{} < p_0^{} < p_2^{}\,$
in Fig.\,\ref{p-velocity-fig},\,\footnote{Note that the equation \eqref{momentum} or \eqref{EOM2}
is invariant under \,$(\phi,\,p)\to (-\phi,\,-p)$.\,
Thus, we have $\,p_1^{}=-p_2^{}\,$ in Fig.\,\ref{p-velocity-fig}.}\,
there exist three $\dot{\phi}$ values obeying Eq.\,(\ref{EOM2}).
Thus, at any moment the propagation of this particle cannot be determined,
and since the switching from one $\,\dot{\phi}\,$ value
to another one could occur instantly,
the usual picture of motion is fully lost.

Hence, for systems with a non-convex Lagrangian such as (\ref{original_L}),
the construction of single-valued Hamiltonian in conjugate momentum space is
challenging. Related issues also arise in cosmology models \cite{Cos1,Cos2,Cos2b},
in extensions of Einstein gravity involving topological invariants \cite{GR-topo,GR-topo2},
and in theories of higher-curvature gravity \cite{GR-new}.
To tackle this, a few approaches were proposed in the literature \cite{PRA,FW-3,nankai}.

As we will show, the real problem with multi-valued Hamiltonian
lies in the conventional Legendre transformation (LT)
which cannot be naively applied to non-convex Lagrangian (\ref{original_L}).
In this work, we propose a new construction of single-valued Hamiltonian
by using Legendre-Fenchel transformation (LFT).
The LFT \cite{convex} is a mathematically rigorous and natural generalization
of the conventional LT for non-convex and non-analytic functions,
although it is not yet widely known to the physics community.
Using this new single-valued Hamiltonian, we will
study the vacuum state and spontaneous breaking of time translation symmetry.
We will further compare the results from different methods, and show that
the LFT is the optimal approach to construct the physical Hamiltonian
for studying non-convex systems.

\section{Legendre-Fenchel Transformation and Construction of Single-Valued Hamiltonian}

The Hamiltonian is usually derived from Lagrangian via
conventional Legendre transformation (LT),
\begin{equation}
H \,=\, p\dot{\phi}-L \,,
\end{equation}
where $\,p\equiv\frac{\partial L}{\partial \dot{\phi}}$\,
is the conjugate momentum.  A prerequisite of the Lengedre transformation is that
the original Lagrangian should be convex and analytic. This is usually
taken for granted. But, for non-convex functions [such as (\ref{original_L})]
or non-analytic case, the LT collapses and its misuse will cause problems, such as
the multi-valuedness of Hamiltonian mentioned above.

To handle the non-convex Lagrangian systems and
consistently derive single-valued Hamiltonians,
the conventional LT is inappropriate and has to be generalized.
Indeed, such a generalized LT is given by the
Legendre-Fenchel transformation (LFT) \cite{convex},
which provides the mathematically rigorous generalization of
the conventional LT, valid for non-convex and/or non-analytic Lagrangian systems.
For the usual case of
convex and analytic functions, the LFT naturally reduces to the LT.
The rigorous LFT method is well-established in mathematics \cite{convex},
but not yet widely known to the physics community.
Applying the general LFT method \cite{convex}, we can rigorously construct
the Hamiltonian,
\begin{equation}
\label{Legendre-Fenchel}
 H(p)\,=  \sup_{\dot{\phi}\in N(p)}
 \!\left[\, p\dot{\phi}-L(\dot{\phi})\,\right] ,
\end{equation}
where the symbol $\,\sup\,$ stands for supremum in mathematics.
In (\ref{Legendre-Fenchel}), $\,N(p)\,$ is defined as
\begin{equation}
\label{domain}
N(p) \,=\,
\left\{\dot{\phi}\,|\sup\, p\dot{\phi}-L(\dot{\phi}) < \infty\right\}
,
\end{equation}
which is required to ensure the finiteness of Hamiltonian for every given $\,p\,$ value.
This provides a physical motivation for generalizing the Legendre transformation.

The LFT construction \eqref{Legendre-Fenchel} is also expected from physics intuition.
Let us consider the steepest-descent approximation or the principle of least action,
which can provide some hints.
This requires the Lagrangian $\,L(\dot{\phi})\,$ to be minimal, and thus
\begin{equation}
H(s,\dot{\phi}) \,=\, s\dot{\phi} - L(\dot{\phi})
\end{equation}
becomes {\it maximal} in $\,\dot{\phi}$\, for every $\,s\,$ value,
where $\,s\,$ is a general quantity independent of $\,\dot{\phi}$\,.\,
This requirement is consistent with the LFT \eqref{Legendre-Fenchel} after replacing
$\,s\,$ by the conjugate momentum $\,p\,$.

To make LFT easy to implement, we present its geometric interpretation.
Let us consider a function,
\begin{equation}
\label{straight-line}
y(\dot{\phi}) \,=\, p(\dot{\phi}-\dot{\phi_0})+L(\dot{\phi_0}) \,,
\end{equation}
which stands for a line passing the point
$\,(\dot{\phi}_0^{},\,L(\dot{\phi}^{}_0))$\,
in the $\,\dot{\phi}-y\,$ plane, with a slope $\,p$\,.\,
Its intercept is
\begin{equation}
Y(p,\dot{\phi}^{}_0) \,=\, -p\dot{\phi}^{}_0 + L(\dot{\phi}^{}_0) \,.
\end{equation}
Thus, we can rewrite Eq.\,(\ref{Legendre-Fenchel}) as
\beqa
H(p) \, &=&
\sup_{\dot{\phi}\in N(p)}\, \left[\, p\dot{\phi}-L(\dot{\phi}) \,\right]
\nn
\\[-2mm]
\\[-2mm]
\nn
\, &=&\sup_{\dot{\phi}\in N(p)}\,-Y(p,\dot{\phi})
\,=\, -\inf_{\dot{\phi}\in N(p)}\,Y(p,\dot{\phi})\,,
\hspace*{9mm}
\eeqa
where the symbol $\,\inf\,$ stands for infimum in mathematics.
Hence, the Hamiltonian $\,H(p)\,$ obtained from LFT just equals the minimal intercept
(with an overall sign flip) of a line with slope $\,p\,$ and crossing
the curve $\,L(\dot{\phi})\,$ in the $\,\dot{\phi}-y\,$ plane.
This approach is also called the supporting line method \cite{convex}.

With this geometric interpretation, we are ready to explicitly construct
the single-valued Hamiltonian via LFT.
For simplicity, we set $\,\kappa=1\,$ from now on. This will not affect
essential features of the analysis and restoring a general parameter
$\,\kappa >0 \,$ is straightforward.
Now, the values of $\,p_1^{}\,$ and $\,p_2^{}\,$ shown in Fig.\,\ref{p-velocity-fig}
are fixed as
\begin{equation}
p_1^{}=-\frac{2}{\,3\sqrt{3}\,} \,,
\hspace*{8mm}
p_2^{}= \frac{2}{\,3\sqrt{3}\,} \,.
\hspace*{5mm}
\end{equation}
Note that the Lagrangian (\ref{original_L}) is analytic. So the minimal intercept
is reached when the line (\ref{straight-line}) is tangent to the Lagrangian curve.
There may be several tangent points for a given slope $\,p\,$,\,
and we should choose the one which minimizes the intercept.
For $\,p \in (\frac{2}{\,3\sqrt{3}},\,+\infty )$,\,
we find from Fig.\,\ref{p-velocity-fig} that the tangent point to the Lagrangian curve
is unique,
\begin{eqnarray}
\label{phi1-p}
\dot{\phi}_1^{}(p) &\!\!=\!\!&
\frac{(\frac{2}{3})^{\frac{1}{3}}}{\,(9p+\sqrt{3}\sqrt{-4+27p^2}\,)^\frac{1}{3}\,}_{}
+\frac{\,(9p+\sqrt{3}\sqrt{-4+27p^2}\,)^\frac{1}{3}_{}\,}
      {2^{\frac{1}{3}}_{}3^{\frac{2}{3}}_{}}\,.
\hspace*{5mm}
\end{eqnarray}
Similar reasoning applies for $\,p\in (-\infty,\,-\frac{2}{\,3\sqrt{3}})$,\,
and the corresponding tangent point is
\begin{eqnarray}
\label{phi3-p}
\dot{\phi}_3^{}(p)\,=
\hspace*{-5mm}
&&-\frac{1+\sqrt{3}\,\mathrm{i}}
  {\,2^{\frac{2}{3}}_{}3^{\frac{1}{3}}_{}(9p+\sqrt{3}\sqrt{-4+27p^2})^{\frac{1}{3}}_{}\,}
\nn\\[-1.5mm]
\\[-1.5mm]
\nn
\hspace*{-5mm}
&&-\frac{\,(1-\sqrt{3}\,\mathrm{i})
   (9p+\sqrt{3}\sqrt{-4+27p^2})^{\frac{1}{3}}\,}{2^{\frac{4}{3}_{}}{3^{\frac{2}{3}}_{}}}.
\hspace*{8mm}
\end{eqnarray}
We note that although $\,\mathrm{i}\,$ appears in (\ref{phi3-p}),
the right-hand-side of (\ref{phi3-p}) is actually real-valued
for $\,p\in (-\infty,\,-\frac{2}{\,3\sqrt{3}})$\,.

\begin{figure}[t]
   \begin{center}
     \includegraphics[width=0.7\textwidth]{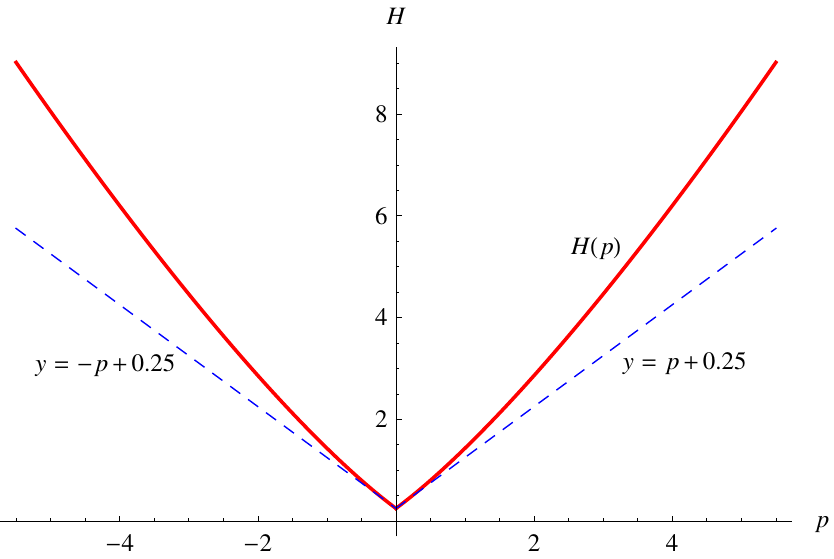}
     \vspace*{-2mm}
   \end{center}
\caption{Hamiltonian $\,H\,$ obtained via Legendre-Fenchel transform (LFT)
as a function of conjugate momentum $\,p\,$.\,  For comparison, we have also added
two extra blue straight dashed-lines $\,y=\pm p+0.25\,$  as reference to make it clear
that our Hamiltonian is not a glue of two straight lines.} 
\label{H-LF-fig}
 \end{figure}

Then, let us consider the interesting range,
$\,p \in \left[-\frac{2}{\,3\sqrt{3}},\,\frac{2}{\,3\sqrt{3}}\right]$.\,
In this case, there are three tangent points with velocities
$\,(\dot{\phi}_3^{},\,\dot{\phi}_2^{},\,\dot{\phi}_1^{})\,$,\,
whose values are shown in Fig.\,\ref{p-velocity-fig}.
To minimize the intercept, we find that for
$\,p \in \left[0,\,\frac{2}{\,3\sqrt{3}}\right]\,$
the right tangent point as shown in (\ref{phi1-p}) gives the minimum;
while for $\,p\in \left[-\frac{2}{\,3\sqrt{3}},\,0\)$\, the left tangent point
as shown in (\ref{phi3-p}) is our choice.
And for $\,p=0\,$,\, the right and left tangent points give the same minimal intercept.
These functions are all real-valued in their defined ranges, and everything is consistent.
With these, we deduce the Hamiltonian from LFT,
%
\beq
\label{H-LF}
\ba{lll}
\hspace*{-5mm}
H(p)\,=\, p\dot{\phi}_1^{}(p)-L(\dot{\phi}_1^{}(p)),\,
& \text{for} & \,p\in[0,\,+\infty)\,,
\\[2mm]
\hspace*{-5mm}
H(p)\,=\, p\dot{\phi}_3^{}(p)-L(\dot{\phi}_3^{}(p)),\,
& \text{for} & \,p\in(-\infty,\,0)\,,
\ea
\eeq
%
where $\,\dot{\phi}_1^{}(p)$\, and $\,\dot{\phi}_3^{}(p)\,$
are defined in Eqs.\,(\ref{phi1-p})-(\ref{phi3-p}).
This Hamiltonian is shown in Fig.\,\ref{H-LF-fig},
where we have added two extra straight dashed-lines for reference
to make it clear that our Hamiltonian is not a combination of
two straight lines as it might appear.  Since we have,
\begin{eqnarray}
H(0) \,=\, 0.25\,, ~~~~
\lim_{p\to 0^\pm}^{} H'(p) \,=\,\pm 1 \,,
\end{eqnarray}
these two dashed-lines are defined as,
$\,H = \pm p \pm 0.25\,$.

\section{Spontaneous Breaking of Time Translation Symmetry and Comparisons}

After constructing the single-valued Hamiltonian via LFT,
we are ready to study its physical application
and compare it with other approaches in the literature, such as
the Hamiltonian path-integral (HPI) method \cite{PRA}
and the naive multi-valued Hamiltonian (\ref{muti-H}).
We will show that our Hamiltonian (\ref{H-LF}) gives the optimal
description of systems with non-convex Lagrangians such as (\ref{original_L}).

 \begin{figure}[t]
   \begin{center}
     \includegraphics[width=0.7\textwidth]{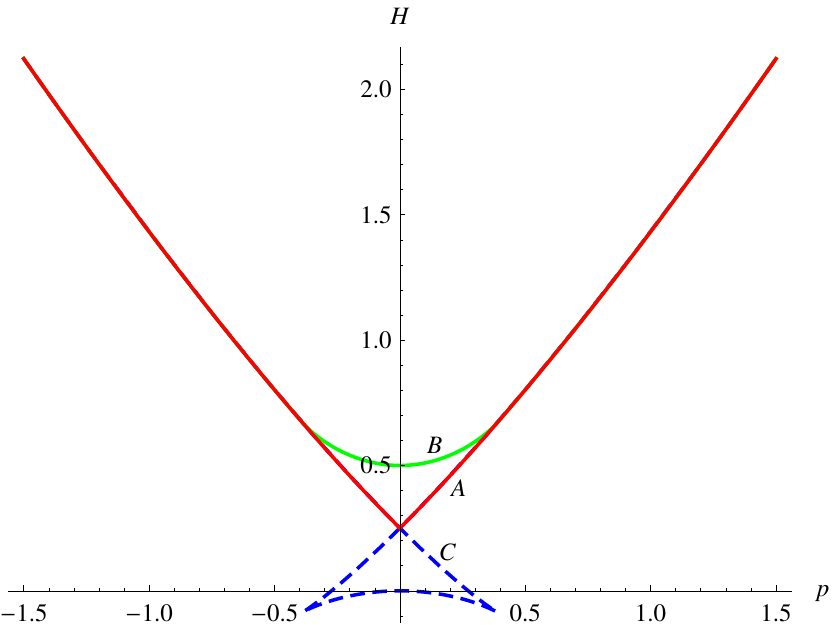}
     \vspace*{-2mm}
   \end{center}
 \caption{Comparison of different Hamiltonians: Curve-A (red) and Curve-B (green) are obtained
 via our LFT approach and HPI method\,\cite{PRA}, respectively;
 Curve-C is the multi-valued Hamiltonian which overlaps with Curve-A and has an extra
 ``swallow tail'' (blue dashed). These three curves coincide over wide ranges except
 the central domain around $\,p=0\,$.}
 \label{H-combine-all-fig}
 \end{figure}

In Fig.\,\ref{H-combine-all-fig},
we present the Hamiltonians derived by the LFT approach (Curve-A),
the HPI method (Curve-B), and the multi-valued Hamiltonian method (Curve-C).
In particular, Curve-A is given by the new Hamiltonian (\ref{H-LF}), and
Curve-C is from the multi-valued Hamiltonian (\ref{muti-H}).
We see that the Hamiltonians of our LFT approach (Curve-A) and the HPI method
(Curve-B) are both single-valued. Furthermore, Curve-A predicts a ground state
of lower energy than that of Curve-B.

Then, we analyze the corresponding Lagrangians by the three methods.
For the LFT approach, we deduce the revised Lagrangian from the single-valued Hamiltonian
(\ref{H-LF}),
\begin{eqnarray}
\label{revised-L}
L_{\text{LFT}}^{}(\dot{\psi}) \,&=&
\frac{1}{4}\dot{\psi}^4-\frac{1}{2}\dot{\psi}^2 ,\,~~~
\text{for}~ \dot{\psi}\in(-\infty,-1) \cup (1,+\infty),
\nn\\[1mm]
L_{\text{LFT}}^{}(\dot{\psi}) \,&=&-\frac{1}{4} \,,\,
\hspace*{15mm}
\text{for}~\dot{\psi} \in [-1,\,1]\,,
\label{L-LFT}
\end{eqnarray}
where $\,\psi\,$ is the coordinate of the configuration space inferred from
the Hamiltonian (\ref{H-LF}), and needs not to match the previous $\,\phi\,$.\,
Eq.\,\eqref{L-LFT} differs from the original Lagrangian (\ref{original_L})
in the range $\,\dot{\psi}\in [-1,\,1]$\,.\,
It also differs from the Lagrangian obtained by the HPI method \cite{PRA}.

\begin{figure}[t]
   \begin{center}
     \includegraphics[width=0.7\textwidth]{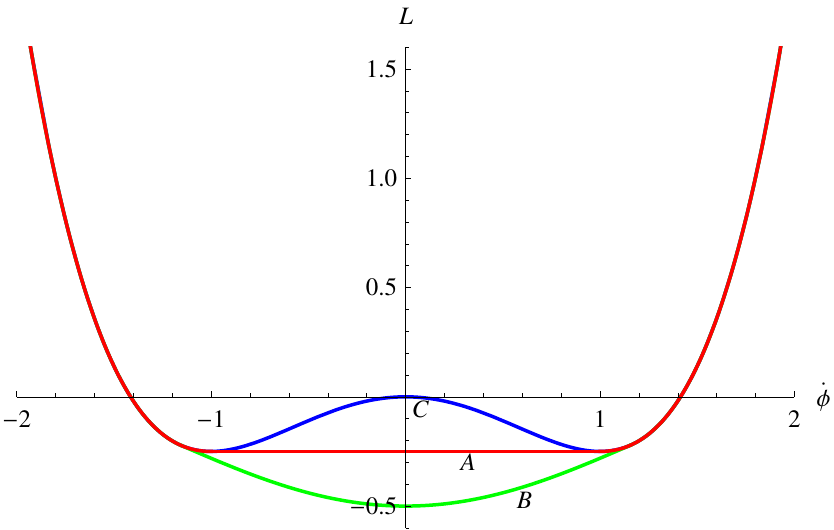}
     \vspace*{-3mm}
   \end{center}
   \caption{Comparison of different Lagrangians: Curve-A and -B are Lagrangians inferred
   from the present LFT Hamiltonian (\ref{H-LF}) and the HPI Hamiltonian \cite{PRA},
   respectively; Curve-C is the original Lagrangian (\ref{original_L}).}
   \label{L-compare-fig}
 \end{figure}

We present the comparison in Fig.\,\ref{L-compare-fig}.
It is apparent that the Lagrangian (\ref{L-LFT}) is
the convex hull of the original Lagrangian (\ref{original_L}),
i.e., the largest convex function satisfying
$\,L_{\text{LFT}}^{}(\dot{\phi})\leqq L(\dot{\phi})$\,.\,
This is expected. We note that the revised Lagrangian \eqref{revised-L}
may be viewed as the ``kinetic version" of Maxwell construction
for thermodynamic free energy \cite{peskin}, which means that the state expressed
as the red straight-line part of Fig.\,\ref{L-compare-fig} is a mixture of
$\,\dot\phi =-1\,$ and $\,\dot\phi=+1$\, states,
similar to the mixing state of water and vapor during evaporation.
This also leads to Eq.\,\eqref{LF-results-new}.

Now, we are ready to discuss the vacuum state and
the spontaneous breaking of the time translation symmetry.
Using the multi-valued Hamiltonian, the analysis in \cite{FW-1}
shows that the vacuum state obtained at the cusp of the Hamiltonian gives,
\begin{equation}
\label{Wilczek-results}
H_0^{} = -\frac{1}{12} \,,~~~~
p_0^{} = \pm\frac{2}{\,3\sqrt{3}\,} \,,~~~~
\dot{\phi}_0^{} = \mp\sqrt{\frac{1}{3}}\,,
\end{equation}
leading to spontaneously broken time translation invariance.
Next, for the HPI method \cite{PRA},
one can infer the vacuum state from Fig.\,\ref{H-combine-all-fig},
\begin{equation}
\label{pra-results1}
H_0^{} =\frac{1}{2}\,, ~~~~
p_0^{}=0 \,.
\end{equation}
Since the HPI does not provide a clear relation
between $\,\dot{\phi}\,$ and $\,p\,$,\,
we may use the canonical equation of the Hamiltonian given by
HPI to obtain the velocity of vacuum state,
\begin{equation}
\label{pra-results2}
\left.
\dot{\psi}_0^{}\,=\, \frac{\partial H(p)}{\partial p}\right|_{p=0}^{} =\, 0\,,
\end{equation}
where $\,\psi\,$ is the coordinate of the configuration space inferred from
the Hamiltonian via usual LT.
Thus, the time translation symmetry is unbroken in the HPI formalism.

Finally, for the LFT approach, we can deduce the vacuum state
from Fig.\,\ref{H-LF-fig} and Eqs.\,(\ref{phi1-p})-(\ref{phi3-p}),
\begin{equation}
\label{LF-results}
H_0^{} \,=\, \frac{1}{4} \,,~~~~
p_0^{} \,=\, 0 \,,~~~~
\dot{\phi}_0^{} \,=\, \pm 1 \,.
\hspace*{8mm}
\end{equation}
In this case, the time translation symmetry is also spontaneously broken.

We also note that it is better to take the original Lagrangian
(\ref{original_L}) only as a starting point
because it is non-convex and thus ill-defined, as commented in \cite{PRA}.
After obtaining a physical Hamiltonian, we can rederive a revised physical Lagrangian,
together with a revised coordinate $\,\psi\,$ of the configuration space, via the LFT.
So, let us derive this coordinate $\,\psi\,$ from the Hamiltonian (\ref{H-LF}),
which needs not to match the previous $\phi\,$.\,
Then, inspecting the vacuum state of Curve-A in Fig.\,\ref{H-combine-all-fig},
we deduce
\begin{equation}
\label{LF-results-new}
\dot{\psi}_0^{} \,\in\, [-1,\,1] \,,
\end{equation}
i.e., the velocity of vacuum state could pick up any value in the range $[-1,1]$.\,
Eq.\,(\ref{LF-results-new}) should precisely describe the vacuum state.
We note that the values of $\,\dot{\psi}_0^{} \in [-1,\,1]$\,
always spontaneously break the time translation symmetry, except a single point at
$\,\dot{\psi}_0^{}=0\,$
which however only has zero measure over the full interval $[-1,1]$.\,

\section{Conclusions and Discussions}

The recent inspiring works of Shapere and Wilczek \cite{FW-3,FW-1}
opened up renewed interests in studying systems with non-convex Lagrangians and
spontaneous breaking of time translation symmetry, as well as applications to
cosmology and gravitation\,\cite{Cos1,Cos2,Cos2b,GR-topo,GR-topo2,GR-new}.
In this work, we proposed a new construction of single-valued Hamiltonian by applying
Legendre-Fenchel transformation (LFT) \cite{convex},
which rigorously generalize the conventional
Legendre transformation for non-convex and non-analytic functions.
We show that this provides a consistent and optimal formulation
in which the corresponding revised Lagrangian is the convex hull of the original Lagrangian.
Then, we studied the vacuum state and the spontaneous breaking of
time translation symmetry, as shown in Fig.\,\ref{H-LF-fig}.
We further compared our predictions with those from other methods in the literature.
The results inferred from different methods are distinctive with each other
(Fig.\,\ref{H-combine-all-fig}), and can be discriminated by experiments.
In Fig.\,\ref{L-compare-fig}, we compared our revised Lagrangian \eqref{L-LFT}
with those from other methods.
Using the new coordinate of configuration space $\,\psi\,$,\,
we inferred the degenerate vacuum states with
(\ref{LF-results-new}).

Finally, it is interesting to further
apply our new LFT approach to related systems, such as specific models of cosmology and gravitation
\cite{Cos1,Cos2,Cos2b,GR-topo,GR-topo2,GR-new}.
For instance, Refs.\,\cite{Cos2,Cos2b} proposed the ghost inflation scenario where an inflationary
de Sitter phase is achieved with a ghost condensate.
It consistently modifies gravity in the infrared, and is realized via a derivatively coupled
ghost scalar field $\,\phi\,$ which forms condensates with non-zero velocity in the background,
$\,\langle\dot{\phi}\rangle = M^2\neq 0\,$.\,
This is a new kind of physical fluid filling the universe and has
fluctuation $\,\hat{\phi}\,$ defined as, $\,\phi = M^2t + \hat{\phi}\,$.\,
This scenario gives an alternative realization of the de Sitter phase,
and the scalar $\,\phi\,$ can naturally serve as inflaton \cite{Cos2b}.
The nonzero ghost condensate $\,\langle\dot{\phi}\rangle \neq 0\,$
also spontaneously breaks time translation symmetry, which is just the physical picture
quantitatively demonstrated in the present work by applying our new LFT method.
Indeed, the LFT method provides a rigorous and consistent way to analyze such cosmological systems.
As another example, the recent work (in the first paper of Ref.\,\cite{GR-new})
studied phase transitions of higher-curvature gravity theories.
Similar to our Eq.\,(\ref{original_L}), it considered an extended non-convex Lagrangian,
$\,L = \fr{1}{2}\dot{\phi}^2 -\fr{1}{3}\dot{\phi}^3 +\fr{1}{17}\dot{\phi}^4\,$\,,\,
by applying the LFT method to derive its convex hull (similar to the Curve-A of
our Fig.\,\ref{L-compare-fig}). The Lovelock gravity was taken as an explicit model
for the analysis\,\cite{GR-new}.  More applications of our LFT method to such gravity and
cosmology models will be pursued further.

\vspace*{6mm}
\noindent
{\bf\large Acknowledgements}\\[2mm]
We thank Nima Arkani-Hamed  for useful discussions.
This work was supported by National NSF of China (under grants 11275101, 11135003)
and National Basic Research Program (under grant 2010CB833000).

\newpage


\noindent
{\bf\large Appendix}

\vspace*{5mm}
\noindent
{\bf A.~Derivations of Eq.\,(18) and Eq.\,(23)}
\vspace*{3mm}

In this Appendix,
we present the detailed derivations of Eq.\,(\ref{revised-L}) and
Eq.\,(\ref{LF-results-new}) from Eq.\,(\ref{H-LF}) under the LFT.
Similar to Eq.\,(\ref{Legendre-Fenchel}), we have
\beqa
\label{revised-L-definition}
L_{\text{LFT}}^{}(\dot{\psi}) \,=\,
\sup_{p\in M(\dot{\psi})} \left[\, p \dot{\psi} - H(p) \,\right] ,
\eeqa
where
\begin{equation}
M(\dot{\psi}) \,=\,
\left\{p\,|\sup\, p \dot{\psi} - H(p) < \infty \right\}
\end{equation}
guarantees the finiteness of the Lagrangian for every given $\,\dot{\psi}$\,.\,
Following the geometric explanation in Sec.\,2,
Eq.\,(\ref{revised-L-definition}) can also be regarded as the minimal intercept
(with an overall sign flip) of a straight line with slope $\,\dot{\psi}\,$ and crossing
the curve $\,H(p)\,$ in the $\,(p,\,L_{\text{LFT}}^{})\,$ plane.
Thus, from Fig.\,\ref{H-LF-fig} and for $\,\dot{\psi}\,\in (1,\infty)$,\,
we obtain the minimal intercept
when the line is tangent to $\,H(p)\,$,\, and the relation between
$\,\dot{\psi}\,$ and $\,p\,$ is,
$\,p={\dot{\psi}}^3-\dot{\psi}\,\in\,(0,\,\infty)$,\,
derived from $\dot{\psi}= {\partial H}/{\partial p}$ in this domain.
The same reasoning applies for $\,\dot{\psi}\,\in (-\infty,-1)$.\,
So, we deduce,
\beq
\label{eq:p-psi-1}
\ba{lll}
\hspace*{-4mm}
p & =\, {\dot{\psi}}^3-\dot{\psi}\,\in\,(0,\,\infty)\,,
  & ~~~\text{for}~~ \dot{\psi}\,\in (1,\infty)\,,
\\[2mm]
\hspace*{-4mm}
p & =\, {\dot{\psi}}^3-\dot{\psi} \,\in\,(-\infty,\,0)\,,
  & ~~~\text{for}~~ \dot{\psi}\,\in (-\infty,-1)\,.
\ea
\eeq
For the region $\,\dot{\psi}\in [-1,\,1]$\,,\,  we reach the minimal intercept
when the line crosses the curve $\,H(p)\,$ at point \,$(0,\,H(0))$.\, Thus, we have
\beqa
\label{eq:p-psi-2}
p\,=\,0\,,\,  ~~~~\text{for}~~ \dot{\psi} \in [-1,\,1]\,,
\eeqa
which just coincides Eq.\,(\ref{LF-results-new}).
Hence, unlike the previous relation (\ref{momentum}),
our Eqs.\,(\ref{eq:p-psi-1})-(\ref{eq:p-psi-2}) give the conjugate-momentum $\,p\,$
as a monotonic function of velocity $\,\dot{\psi}\,$.\,
With the above, we can derive the new Lagrangian from Eq.\,(\ref{revised-L-definition}),
\beqa
\label{revised-L-2}
\hspace*{-3mm}
L_{\text{LFT}}(\dot{\psi}) &\!=\!&
\frac{1}{4}\dot{\psi}^4-\frac{1}{2}\dot{\psi}^2 \,,
~~ \hspace*{3mm}  \text{for}~ \dot{\psi}\,\in (-\infty,-1)\cup (1,\infty).
\non
\\
\hspace*{-3mm}
L_{\text{LFT}}^{}(\dot{\psi}) &\!=\!& -H(0) = -\frac{1}{4}\,,
~~ \text{for}~ \dot{\psi} \in [-1,\,1].
\eeqa
This just reproduces our Eq.\,(\ref{revised-L}).

\vspace*{6mm}
\noindent
{\bf B. Discussing the Previous Approaches}
\vspace*{3mm}

In this Appendix, for completeness we discuss the previous different attempts
in the literature \cite{PRA,FW-3}
for tackling the multi-valued Hamiltonian, which are independent of
our current study.
Ref.\,\cite{PRA} adopted a Hamiltonian path-integral (HPI) method,
defined in the position-velocity space, where the transition amplitude
$\,\left<\phi_2^{},t_2^{}|\phi_1^{},t_1^{}\right>\,$ is given by
%
\begin{eqnarray}
\label{transition-amplitude}
\int\!\!\!\!
\int\!\!\mathcal{D}\phi (t)\mathcal{D}u(t)\prod_t\frac{\partial^2 L}{\partial u\partial u} \exp\!\left\{\frac{i}{\hbar}S_H^{}[\phi(t),u(t)]\right\} .
\hspace*{12mm}
\end{eqnarray}
Then, the authors tried to sum up all paths in $(\phi,\,u)$ which correspond
to the same path in phase-space $(\phi,\,p)$.\,  This would result in an
effective Hamiltonian $H_{\text{eff}}^{}$ as a single-valued function
of conjugate-momentum $p\,$.\,
Note that the Hamiltonian $H$ is conventionally defined
in the canonical space $(\phi,\,p)$ via usual Legendre transformation (LT),
which is multi-valued function of velocity $\,u\,$ (or, $\dot{\phi}$)
for each given $\,p \in (p_1^{},\,p_2^{})\,$.\,
So, given such an ill-defined Hamiltonian, it is fully unknown {\it a priori}
which solutions of $\dot{\phi}$ or which paths in $(\phi,\,u)$ space are physically
(un)acceptable. As shown in Sec.\,2, because the conventional LT is ill-defined
for non-convex Lagrangian (\ref{original_L}), it is likely that some or all of the
$\,\dot{\phi}\,$ solutions are nonphysical. Hence, the naive sum of all paths
in $(\phi,\,u)$ \cite{PRA} may not be physically meaningful,
although one could take it as a working ansatz \cite{PRA}.

The recent approach \cite{FW-3} proposed a new method
of branched quantization, which unfolds the non-monotonic
$\,p(\dot{\phi})\,$ by redefining the conjugate momentum as,
%
\beq
\label{xi-p}
\ba{ll}
\xi \,\equiv\, p-p_2^{}+p_1^{},  & ~~~\text{for}~~~ \xi \leqq p_1^{}\,,
\\[1.5mm]
\xi \,\equiv\, -p+p_2^{}+p_1^{}, & ~~~\text{for}~~~ p_1^{} \leqq \xi \leqq p_2^{} \,,
\\[1.5mm]
\xi \,\equiv\, p+p_2^{}-p_1^{},  & ~~~\text{for}~~~ p_2^{} \leqq \xi \,.
\ea
\eeq
%
We can reexpress (\ref{xi-p}) as follows,
%
\beq
\label{xi-p-problem}
\ba{ll}
\xi \,\equiv\, p-p_2^{}+p_1^{},  & ~~~\text{for}~~~ p \leqq p_2^{}\,,
\\[1.5mm]
\xi \,\equiv\, -p+p_2^{}+p_1^{}, & ~~~\text{for}~~~ p_1^{} \leqq p \leqq p_2^{}\,,
\\[1.5mm]
\xi \,\equiv\, p+p_2^{}-p_1^{},  & ~~~\text{for}~~~ p_1^{} \leqq p \,.
\ea
\eeq
%
As clearly shown, the ranges of $\,p\,$ in the three branches overlap with each other
in (\ref{xi-p-problem}). So, each given $\,p\,$ corresponds to three
$\,\xi\,$ values over the range of $\,p_2^{}> p > p_1^{}=-p_2^{}\,$.\,
As such, the Hamiltonian is still a multi-valued function of new conjugate momentum
$\,\xi\,$, because it is easy to verify from (\ref{momentum}) or (\ref{EOM2})
that the equation $\,\dot{\phi}^3 -\kappa\dot{\phi} = p = -\xi\,$  still has
three solutions $\,\dot{\phi}\,$ for each given $\,\xi\in (p_1^{},\,p_2^{})\,$
in \eqref{xi-p}.

 \begin{figure} 
   \begin{center}
     \includegraphics[width=0.7\textwidth]{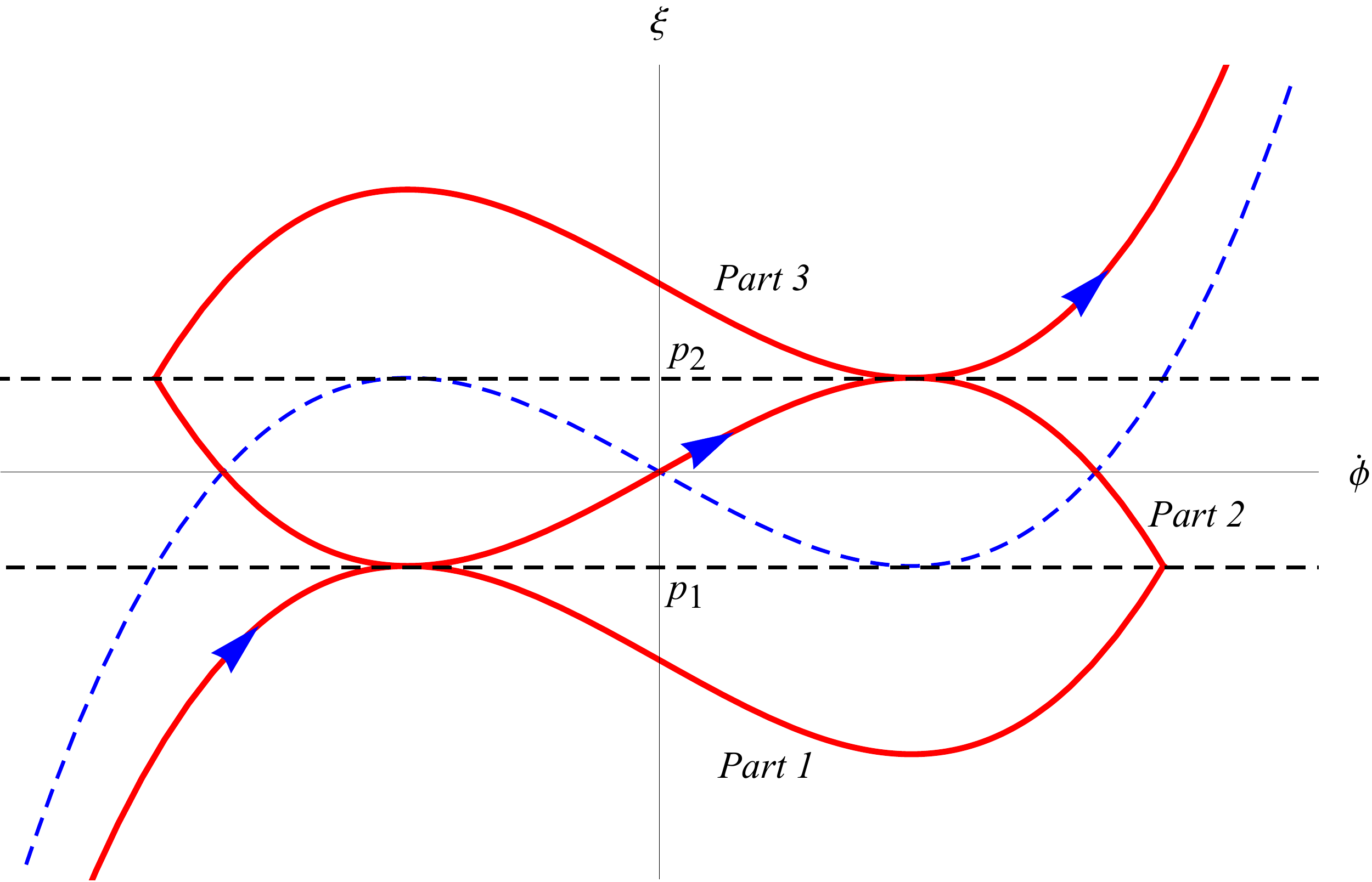}
     \vspace*{-3mm}
   \end{center}
  \caption{The relation between $\,\dot{\phi}\,$ and $\,\xi\,$,\,
  inferred from Eq.\,(\ref{xi-p}) or (\ref{xi-p-problem}).
  Part-1, -2 and -3 of the red solid curve is obtained by the first, second and third branch of
  Eq.\,(\ref{xi-p}) or (\ref{xi-p-problem}), respectively.
  The blue dashed curve $\,p(\dot{\phi})\,$ is re-plotted from Fig.\,\ref{p-velocity-fig}
  as a reference, with the vertical axis labeled by $\,p\,$.}
 \label{xi-p-wrong}
 \end{figure}

To be intuitive, we further explicitly present
Eq.\,(\ref{xi-p}) or Eq.\,(\ref{xi-p-problem}) in Fig.\,\ref{xi-p-wrong}
as denoted by the red solid curve.
Part-1, -2, and -3 of the red curve are obtained by the first, second,
and third branch of Eq.\,(\ref{xi-p}) or (\ref{xi-p-problem}), respectively.
For reference, we also plot the blue dashed curve $\,p(\dot{\phi})$\,
(from Fig.\,\ref{p-velocity-fig}) with the vertical axis labeled by $\,p\,$.\,
From the red curve in Fig.\,\ref{xi-p-wrong}, it is apparent
that $\,\xi\,$ is not a simple function of $\,\dot{\phi}\,$.\,
There are three routes which form two extra loops;
and only the middle one (marked by arrows) is what Ref.\,\cite{FW-3} hoped to realize.
But, the other two routes are also contained in Eq.\,(\ref{xi-p}) or (\ref{xi-p-problem}).
Thus, Eqs.\,(\ref{xi-p})-(\ref{xi-p-problem}) appear ill-defined and
the problem of multi-valued Hamiltonian remains unsolved.
If we really want to pick up one route out of the three options,
say, the middle one as implied in Ref.\,\cite{FW-3},
we have to add further restrictions to realize it,
such as by imposing the supremum condition in the LFT (\ref{Legendre-Fenchel}).
Since the LFT is both physically well-motivated and mathematically rigorous,
it is the optimal approach to handle such singular systems.

In addition, Ref.\,\cite{FW-3} tried to impose certain boundary conditions
for ensuring the unitary evolution for multivalued Hamiltonian.
As we explicitly clarify in the following, unfortunately this does not work.
Ref.\,\cite{FW-3} defined three branches of wave functions in the paragraph below its
eq.\,(3), i.e.,
$\,\psi_1^{}(p)\,$ for $\,-\infty< p \leqq p_+^{}$\,,\,
$\,\psi_2^{}(p)\,$ for $\,p_-^{} \leqq p \leqq p_+^{}\,$,\, and
$\,\psi_3^{}(p)\,$ for $\,p_-^{} \leqq p <\infty\,$.\,
It is clear that all three components cover the range $\,p_-^{} \leqq p \leqq p_+^{}\,$,\,
as stated at the end of this paragraph\,\cite{FW-3}.
Hence, for the boundary conditions (6) of Ref.\,\cite{FW-3}
to ensure the vanishing probability current $\,j=0\,$
at the two junctions $\,p=p_\pm^{}$,\, all the three branches of wave functions
$\,\psi_{1,2,3}^{}(p)\,$ [rather than just two branches $\psi_{1,2}^{}(p)$]
will contribute to $\,j\,$,\, so they have to be included together.
This means that the eq.\,(6) of Ref.\,\cite{FW-3} with $\psi_{1,2}^{}(p)$ alone
cannot actually ensure $\,j=0\,$
at $\,p=p_+^{}\,$,\, and thus the integrated probability $\int\!\rho$\,
is not conserved (unlike the claim of Ref.\,\cite{FW-3});
the same flaw exists for $\,p=p_-^{}$.\,  Furthermore,
because of the $\,\psi_3^{}(p)\,$ contributions to the boundary conditions at $\,p=p_\pm^{}$,\,
eq.\,(6) (at $\,p=p_+^{}\,$) and its analogues (at $\,p=p_-^{}\,$)
will contain {\it extra} $\,4\,(=2+2)\,$ constraints,
so there are actually $\,4+4=8\,$ constraints at $\,p=p_\pm^{}$.\, Together with the two normalizability
conditions at $\,p\to\pm\infty\,$ \cite{FW-3}, the total number of boundary constraints is 10\,,\,
which does not match the number of integral constants \,$n=2\times 3 =6$\,
of the Schr\"{o}dinger equations for $\,\psi_{1,2,3}^{}(p)$.\,
Unfortunately, these complete boundary conditions over-constrain the system,
so this boundary condition method\,\cite{FW-3} does not really work.
(The similar problems also exist for related applications \cite{SWX2012}.)


\begin{thebibliography}{99}


\bibitem{FW-3}
Alfred Shapere and Frank Wilczek,
Phys.\ Rev.\ Lett.\ {\bf 109} (2012) 200402
[arXiv:1207.2677 [quant-ph]].


\bibitem{FW-1}
Alfred Shapere and Frank Wilczek, 
Phys.\ Rev.\ Lett.\ {\bf 109} (2012) 160402
[arXiv:1202.2537 [cond-mat]].


\bibitem{PRA}
M.\ Henneaux, C.\ Teitelboim, and J.\ Zanelli,
Phys.\ Rev.\ A\ {\bf 36} (1987) 4417.


\bibitem{Cos1}
C.\ Armendariz-Picon, T.\ Damour and V.\ F.\ Mukhanov,
Phys.\ Lett.\ B\ {\bf 458} (1999) 209 [arXiv:hep-th/9904075].


\bibitem{Cos2}
N.\ Arkani-Hamed, H.\,C.\ Cheng, M.\ Luty, and S.\ Mukohyama,
JHEP {\bf 0405} (2004) 074 [arXiv:hep-th/0312099].


\bibitem{Cos2b}
N.\ Arkani-Hamed, P.\ Creminelli, S.\ Mukohyama, and M.\ Zaldarriaga,
JCAP {\bf 0404} (2004) 001 [arXiv:hep-th/0312100].


\bibitem{GR-topo}
C. Teitelboim and J. Zanelli,
Class.\ Quant.\ Grav.\ {\bf 4} (1987) L125.



\bibitem{GR-topo2}
C. Teitelboim and J. Zanelli,
in \emph{Constraints Theory and Relativistic Dynamics},
eds.\ C.\ Longhi and L.\ Lusanna
(World-Scientific, 1987).


\bibitem{GR-new}
For the latest developments,
X.\ O.\ Camanho, J.\ D.\ Edelstein, G.\ Giribet, and A.\ Gomberoff,
arXiv:1311.6768 [hep-th];
E.\ Avraham and R.\ Brustein,
arXiv:1401.4921 [hep-th];  and references therein.


\bibitem{nankai}
L.\ Zhao, P.\ Yu, and W.\ Xu,
Mod.\ Phys.\ Lett.\ A\,{\bf 28} (2013) 1350002 [arXiv:1206.2983 [hep-th]].



\bibitem{convex}
For a comprehensive monograph, see:
R.\ T.\ Rockafellar,
\emph{Convex Analysis}, Princeton\ University\ Press,\ Princeton,\ 1970.


\bibitem{peskin}
E.g., M.\ E.\ Peskin and D.\ V.\ Schroeder,
``An Introduction to Quantum Field Theory", (cf.\ p.368-369),
Westview Press, 1995.


\bibitem{SWX2012}
A.\ D.\ Shapere, F.\ Wilczek, and Z.\ Xiong,
arXiv:1210.3545 [hep-th].

\end{thebibliography}

\vspace*{6mm}

\end{document}